\documentclass[twocolumn,showpacs,preprintnumbers,superscriptaddress,amsmath,floatfix,amssymb,prd]{revtex4}
\usepackage[colorlinks=true]{hyperref}
\usepackage{graphicx}

\newcommand{\comment}[1]{}

\newcommand{\lr}[1]{ \left( #1 \right) }

\newcommand{\tr}{ {\rm Tr} \, }

\newcommand{\ket}[1]{ \, | #1 \rangle }
\newcommand{\bra}[1]{ \langle #1 | \, }
\newcommand{\braket}[2]{ \langle #1 | \, #2 \, | #1 \rangle }

\newcommand{\expa}[1]{ \exp{\left( #1 \right)} }

\newcommand{\logo}{\\ \vskip -18mm
\leftline{\includegraphics[scale=0.3,clip=false]{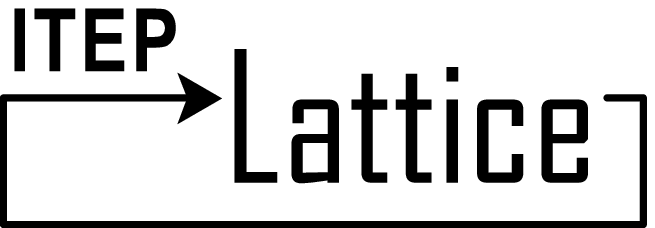}} \vskip 10mm}

\begin{document}
\sloppy
\preprint{ITEP-LAT/2009-03}

\title{On the dynamics of large-$N$ $O\lr{N}$-symmetric quantum systems at finite temperature\logo}
\author{P. V. Buividovich}
\email{gbuividovich@gmail.com}
\affiliation{JIPNR, National Academy of Science, 220109 Belarus, Minsk, Acad. Krasin str. 99}
\affiliation{ITEP, 117218 Russia, Moscow, B. Cheremushkinskaya str. 25}
\date{March 25, 2009}
\begin{abstract}
  Time evolution of a perturbed thermal state is studied in a quantum-mechanical system with $O\lr{N}$ symmetry. In the limit of large $N$, time dependence of $O\lr{N}$-singlet expectation values can be described by classical equations of motion in a one-dimensional potential well. Time dependence of the perturbation is then described by a linear differential equation with time-dependent periodic coefficient. This equation, depending on the parameters, admits either exponentially growing/decaying or periodically oscillating solutions. It is demonstrated that only the latter possibility is actually realized, thus in such a system there is no redistribution of initial perturbation over all $N$ degrees of freedom.
\end{abstract}
\maketitle

 This work was motivated by several closely related nonstationary problems in quantum field theory. The first two are Schwinger pair creation in a strong electric field \cite{Schwinger:51:1} and decay of cosmological constant in de Sitter space \cite{Polyakov:08:1, Woodard:96:1, Buividovich:08:4}. Another problem is ``fast scrambling'' of information in matrix quantum mechanics considered recently in \cite{Susskind:08:1, Polchinski:08:1}. In all cases initial state of the system is classically stable but decays due to quantum effects. Typically, only a few degrees of freedom out of some large or infinite number are excited initially, but due to interactions the energy and information content of the initial state is redistributed more or less equally among all degrees of freedom.

 In the context of such problems, one is usually interested in expectation values of the form $\bra{in} e^{-i \hat{H} t} \hat{O} e^{i \hat{H} t}  \ket{in}$, where $\hat{H}$ is the Hamiltonian and $\hat{O}$ is some quantity which characterizes a perturbation of the system (for example, field strength for the decay of strong electric field). Practical calculations are usually done either using the perturbation theory, semiclassical approximation or numerical methods. It is clear that perturbative expansion can only lead to the results of the form $\bra{in} e^{-i \lr{\hat{H}_{0} + g \hat{H}_{I}} t} \hat{O} e^{i \lr{\hat{H}_{0} + g \hat{H}_{I}} t}  \ket{in} = \bra{in} \hat{O} \ket{in} + a_{1} g t + a_{2} g^{2} t^{2} + ...$, which are valid either for small $t$ or $g$. It is not surprising that such a dependence on time was indeed obtained, for example, in perturbative calculations of graviton loops in de Sitter background \cite{Woodard:96:1}. On the other hand, semiclassical calculations similar to that of Schwinger \cite{Schwinger:51:1} usually fail to take into account back-reaction processes.

 The aim of this Letter is to obtain an exact description of the time evolution of a perturbed thermal state in a model system with very large number $N$ of interacting degrees of freedom. The Hamiltonian of this model system is $O\lr{N}$-symmetric and has the following form \cite{MakeenkoGaugeMethods, Feinberg:95:1, Fateev:81:1}:
\begin{eqnarray}
\label{Hamiltonian}
 \hat{H} = \sum \limits_{a = 1}^{N} \frac{\hat{\pi}_{a}^{2}}{2} + N V\lr{ \frac{1}{N}\, \sum \limits_{a = 1}^{N} \hat{\phi}_{a}^{2} }, 
\end{eqnarray}
where $\hat{\phi}_{a}$, $\hat{\pi}_{a}$ are the coordinate and momentum operators and $V\lr{x}$ grows monotonously for $x > 0$ and does not depend on $N$. The terms in $V\lr{x}$ which are proportional to $x^{k}$ correspond to interactions involving $k$ degrees of freedom. Note that each degree of freedom interacts with each other degree of freedom with equal strength. In the limit of infinite $N$ such system can be thought of as a sort of mean-field approximation to a quantum theory of scalar field. Indeed, if one considers scalar fields on $N$-dimensional lattice, each lattice site interacts equally with $N$ neighbors, and the limit $N \rightarrow \infty$ leads to the standard mean-field approximation. This analogy closely follows the discussion of ``fast scrambling'' in \cite{Susskind:08:1}, where the following problem was considered: suppose one has a system with a large number $N$ of degrees of freedom. How should these degrees of freedom interact in order to redistribute perturbations of a small number of degrees of freedom over all the system most efficiently and quickly? The answer suggested in \cite{Susskind:08:1} is precisely that each degree of freedom should interact with each other degree of freedom with equal strength. Thus one might hope that the system described by the Hamiltonian (\ref{Hamiltonian}) is a ``fast scrambler''.

 Here the following particular problem will be considered: suppose the system (\ref{Hamiltonian}) is initially in a state of thermal equilibrium with density matrix $\hat{\rho}_{0} = \mathcal{Z}^{-1} \, \expa{ - \beta \hat{H}}$, where $\beta$ is the inverse temperature. Then the system is perturbed in such a way that the density matrix $\rho_{0}\lr{\phi_{a}, \phi_{a}'}$ in the coordinate representation is shifted by a constant vector $\xi_{a}$: $\rho_{in}\lr{\phi_{a}, \phi_{a}'} = \rho_{0}\lr{\phi_{a} - \xi_{a}, \phi_{a}' - \xi_{a}}$. $\frac{1}{N} \sum \limits_{a = 1}^{N} \xi_{a}^{2}$ is assumed to be finite in the limit $N \rightarrow \infty$. After that the system is allowed to evolve freely. What will be the time dependence of the expectation value $\phi_{a}\lr{t} = \tr\lr{ \hat{\rho}_{in} \hat{\phi}_{a}\lr{t} } $, where $\hat{\phi}_{a}\lr{t}$ is the coordinate operator in the Heisenberg representation? Will it decay with time because the energy of an initial perturbation is redistributed over all $N$ degrees of freedom, or will it oscillate periodically? The large-$N$ analysis presented here suggests that only the latter possibility is actually realized for any potential $V\lr{x}$.

 Heisenberg equations of motion for $\hat{\phi}_{a}$ and $\hat{\pi}_{a}$ are $\frac{d}{dt}\, \hat{\phi}_{a} = \hat{\pi}_{a}$, $\frac{d}{dt}\, \hat{\pi}_{a} = - 2 V'\lr{\hat{x}} \hat{\phi}_{a}$, where $\hat{x}\lr{t} = \frac{1}{N}\, \sum \limits_{a = 1}^{N} \hat{\phi}_{a}^{2}\lr{t}$ and $V'\lr{x}$ is the derivative of the potential over $x$. Differentiating once again over time, one obtains:
\begin{eqnarray}
\label{second_order_eqm_phi}
\frac{d^{2}}{dt^{2}}\, \hat{\phi}_{a} = - 2 V'\lr{\hat{x}} \hat{\phi}_{a}
\\
\label{second_order_eqm_x}
\frac{d^{2}}{dt^{2}}\, \hat{x} = \frac{2}{N}\, \sum \limits_{a = 1}^{N} \hat{\pi}_{a}^{2} - 4 V'\lr{\hat{x}} \hat{x}
= \nonumber \\ =
\frac{4}{N}\, \hat{H} - 4 V\lr{\hat{x}} - 4 \hat{x} V'\lr{\hat{x}}, 
\end{eqnarray}
where in the last line the operator $\sum \limits_{a = 1}^{N} \hat{\pi}_{a}^{2}$ was expressed in terms of the Hamiltonian (\ref{Hamiltonian}).

 A crucial step now is to take the expectation values of the r.h.s. and l.h.s. of (\ref{second_order_eqm_phi}) and (\ref{second_order_eqm_x}) over the initial state with the density matrix $\hat{\rho}_{in}$. For the ground state of the Hamiltonian (\ref{Hamiltonian}), or, more generally, for the thermal state $\hat{\rho}_{0}$ in the limit of large $N$, expectation values of multiple powers of $\hat{x}$ factorize \cite{MakeenkoGaugeMethods}: $\lim \limits_{N \rightarrow \infty} \tr\lr{\hat{\rho}_{0} \, \hat{x}^{k}} = \tr\lr{\hat{\rho}_{0} \, \hat{x}}^{k}$. Since the density matrix $\hat{\rho}_{in}$ was obtained from $\hat{\rho}_{0}$ by a shift of the coordinates $\phi_{a}$, it is easy to show that the same property holds also for $\hat{\rho}_{in}$: $\lim \limits_{N \rightarrow \infty} \tr\lr{\hat{\rho}_{in} \, \hat{x}^{k}} = \tr\lr{\hat{\rho}_{in} \, \hat{x}}^{k}$. Factorization property is usually considered only for ground states of large-$N$ theories \cite{MakeenkoGaugeMethods, Fateev:81:1}, however, it is easy to show that it holds also for the initial thermal state $\hat{\rho}_{0}$ and for the perturbed state at $t > 0$. To this end one can consider the path integral for $\tr\lr{\hat{\rho}_{in} \hat{\phi}_{a}\lr{t}} = \tr\lr{ \hat{\phi}_{a}\lr{0} \expa{ i \int d\tau\lr{ \hat{H} + J_{a}\lr{\tau} \hat{\phi}_{a} }}}$ and note that it has a saddle point which dominates in the large-$N$ limit. The existence of such a saddle point in the path integral corresponds to factorization property in the canonical formalism. Here the integral goes over the $T$-shaped contour in the complex plane which consists of the double line of length $t$ on the real axis and of the line of length $\beta$ on the imaginary axis. At the ends of this line the usual periodic boundary conditions are imposed. At the intersection of the two lines, a singular source $J_{a}\lr{\tau} = - \xi_{a} \, \frac{d}{d \tau} \, \delta\lr{\tau, 0}$ is inserted into the path integral in order to reproduce the initial perturbation. Path integral over the imaginary axis yields then the thermal ground state and path integral over real time describes evolution of the perturbed state in time. However, the equations of motion (\ref{second_order_eqm_phi}) and (\ref{second_order_eqm_x}) are most easily analyzed in the canonical formalism.

 Assuming factorization,  introducing the expectation values $\phi_{a}\lr{t} = \tr\lr{ \hat{\rho}_{in} \hat{\phi}_{a}\lr{t} }$, $x\lr{t} = \tr\lr{ \hat{\rho}_{in} \hat{x}\lr{t} }$, and taking into account that the Hamiltonian (\ref{Hamiltonian}) does not depend on time, one can obtain a closed system of equations for $\phi_{a}\lr{t}$, $x\lr{t}$:
\begin{eqnarray}
\label{second_order_eqm_phi_ev}
\frac{d^{2} \phi_{a}}{dt^{2}}\,  = - 2 \phi_{a} V'\lr{x}
\\
\label{second_order_eqm_x_ev}
\frac{d^{2} x}{dt^{2}} = - \frac{d}{dx} \, \tilde{V}\lr{x}
\\
\label{x_eff_potential}
\tilde{V}\lr{x} = 4 x \, \lr{ V\lr{x} - \frac{1}{N}\, \tr\lr{\hat{\rho}_{in} \hat{H}} }.
\end{eqnarray}
Initial conditions for the equations (\ref{second_order_eqm_phi_ev}) and (\ref{second_order_eqm_x_ev}) are:
\begin{eqnarray}
\label{initial_cond}
 \phi_{a}\lr{0} = \xi_{a}, \quad \frac{d \phi_{a}}{dt}\lr{0} = 0,
\nonumber \\
 x\lr{0} = x_{0} + \frac{\xi_{a}^{2}}{N}, \quad \frac{d x}{dt}\lr{0} = 0
\end{eqnarray}
where $x_{0} = \tr\lr{\hat{\rho}_{0} \hat{x}}$ is the expectation value of $x$ in the thermal state $\hat{\rho}_{0}$. The last initial condition can be justified in the following way. For any quantum state $\ket{\Psi}$ one has $\braket{\Psi}{ \frac{d}{dt}\, \hat{x} } = \frac{1}{N} \sum \limits_{a = 1}^{N} \braket{\Psi}{ \hat{\phi}_{a} \hat{\pi}_{a} + \hat{\pi}_{a} \hat{\phi}_{a} } =  \frac{i}{N} \sum \limits_{a = 1}^{N} \int d^{N} \phi_{a} \lr{ \bar{\Psi} \phi_{a} \frac{\partial}{\partial \phi_{a}} \Psi + \bar{\Psi} \frac{\partial}{\partial \phi_{a}} \lr{\phi_{a} \Psi} } =  \frac{i}{N} \sum \limits_{a = 1}^{N} \int d^{N} \phi_{a} \lr{ \bar{\Psi} \phi_{a} \frac{\partial}{\partial \phi_{a}} \Psi - \lr{\phi_{a} \Psi} \frac{\partial}{\partial \phi_{a}} \bar{\Psi}  } $. The expectation value $\braket{\Psi}{ \frac{d}{dt}\, \hat{x} }$ is equal to the sum of the above expression over all eigenstates $\ket{\Psi_{n}}$ of the Hamiltonian (\ref{Hamiltonian}). The states with the wave functions $\Psi_{n}$ and $\bar{\Psi}_{n}$ are both the eigenstates of the Hamiltonian (\ref{Hamiltonian}) with the same energy, and therefore both of them enter the thermal density matrix with equal weights. This means that for each pair $\Psi_{n}$ and $\bar{\Psi}_{n}$ the terms in the brackets give zero in total. Therefore $\frac{dx}{dt}\lr{0} = \tr\lr{ \hat{\rho}_{in}  \frac{d}{dt}\, \hat{x} } = 0$.

\begin{figure}
  \includegraphics[width=6cm]{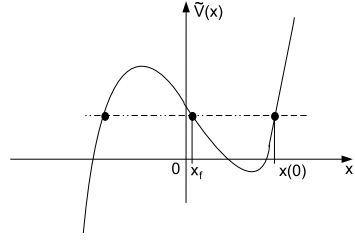}\\
  \caption{A typical shape of the potential $\tilde{V}\lr{x}$.}
  \label{fig:sigma_potential}
\end{figure}

 Consider now the potential $\tilde{V}\lr{x}$. It depends on the initial conditions due to the term $\tr\lr{\hat{\rho}_{in} \hat{H}} = \tr\lr{\hat{\rho}_{in} \frac{\hat{\pi}^{2}_{a}}{2}} + N V\lr{x\lr{0}}$, which can be expressed in terms of $x_{0}$ and $\xi_{a}$. To this end note that $\tr\lr{\hat{\rho}_{in} \frac{\hat{\pi}^{2}_{a}}{2}} = \tr\lr{\hat{\rho}_{0} \frac{\hat{\pi}^{2}_{a}}{2}}$ and make use of the virial theorem, which states that $\tr\lr{ \hat{\rho}_{0} \frac{\hat{\pi}_{a}^{2}}{2}} = \tr\lr{ \hat{\rho}_{0} \frac{1}{2}\, \hat{\phi}_{a} \frac{\partial \hat{H}}{\partial \phi_{a}} } = N x_{0} V\lr{x_{0}}$. Thus the potential $\tilde{V}\lr{x}$ is:
\begin{eqnarray}
\label{x_potential}
\tilde{V}\lr{x} = 4 x V\lr{x} - 4 x \lr{x_{0} V'\lr{x_{0}} + V\lr{x\lr{0}}}
\end{eqnarray}
It is easy to show that if $V\lr{x}$ is a monotonic continuous function for $x > 0$, equation (\ref{second_order_eqm_x_ev}) always describes a finite periodic motion in a one-dimensional potential well. Namely, there is always such $x_{f} < x\lr{0}$, $x_{f} > 0$ that $\tilde{V}\lr{x_{f}} = \tilde{V}\lr{x\lr{0}}$ and $\tilde{V}\lr{x} < \tilde{V}\lr{x\lr{0}}$ for $x_{f} < x < x\lr{0}$, so that $x\lr{t}$ moves periodically between $x_{f}$ and $x\lr{0}$ with some period $T$. A typical shape of $\tilde{V}\lr{x}$ is illustrated on Fig. \ref{fig:sigma_potential}. Note that $\tilde{V}\lr{x}$ is not bounded from below, but depends on $x\lr{0}$ in such a way that the motion of $x$ is always finite and is always restricted to $x > 0$.

 The equations (\ref{second_order_eqm_phi_ev}), (\ref{second_order_eqm_x_ev}) together with the initial conditions (\ref{initial_cond}) are the main result of this work. Thus in the large $N$ limit $x$ moves classically in the  one-dimensional potential well $\tilde{V}\lr{x}$, while $\phi_{a}\lr{t}$ is a solution of the linear differential equation with time-dependent coefficient $2 V'\lr{x}$. Since $x\lr{t}$ is periodic, one can readily apply the Floquet theorem, which states that the equation (\ref{second_order_eqm_phi_ev}) with periodic $x\lr{t}$ can have either two independent oscillating solutions or two solutions one of which grows and the other decays exponentially. In the first case the eigenvalues of the Floquet matrix have unit absolute values and are complex conjugate to each other: $|\lambda_{1}| = |\lambda_{2}| = 1$, $\lambda_{1} = \bar{\lambda}_{2}$. In the other case, which corresponds to parametric resonance between $x$ and $\phi_{a}$, the eigenvalues $\lambda_{1}$, $\lambda_{2}$ are real and $\lambda_{1} \lambda_{2} = 1$.

 One might therefore ask, whether by tuning the parameters of the system it is possible to achieve such a parametric resonance between $x\lr{t}$ and $\phi_{a}\lr{t}$? The answer to this question seems to be negative. Exponentially growing $\phi_{a}\lr{t}$ is clearly prohibited on general physical grounds, since in this case any small perturbation causes the energy of all $N$ degrees of freedom to be collected into one mode. While such solutions may be in principle relevant for spontaneous symmetry breaking, they are very unlikely for monotonic potentials $V\lr{x}$. I was not able to find a general proof of this fact, however, in numerical investigations of the equations (\ref{second_order_eqm_phi_ev}), (\ref{second_order_eqm_x_ev}) for several different $V\lr{x}$ no regions of parametric resonance were found.

\begin{figure}
  \includegraphics[width=6cm]{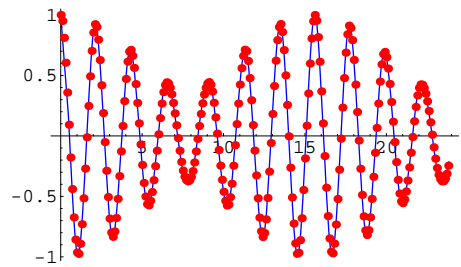}\\
  \caption{The solution $\phi_{a}\lr{t}$ of (\ref{second_order_eqm_phi_ev}) (projected on the direction of $\phi_{a}\lr{0}$) for the potential $V\lr{x} = \frac{w^{2} x}{2} + \frac{\lambda x^{2}}{4}$ at the inverse temperature $\beta = 0.5$. Dots denote the numerical result for $\phi_{a}\lr{t}$ obtained by solving the saddle-point equations for $\phi_{a}\lr{t}$ and $x\lr{t}$ in the path integral representation.}
  \label{fig:phi_vs_t}
\end{figure}

 But probably there can be isolated points in parameter space, for which $\phi_{a}\lr{t}$ decays exponentially? At first sight on equations (\ref{second_order_eqm_x_ev}), (\ref{second_order_eqm_phi_ev}) and (\ref{initial_cond}), it seems that such possibility may indeed be realized, and the system (\ref{Hamiltonian}) may indeed ``scramble'' an initial perturbation. However, this turns out to be impossible as well. The reason for that is, at the bottom of the fact, the invariance of the system (\ref{Hamiltonian}) under time reflection. Indeed, $\phi_{a}\lr{t}$ can only decay exponentially if the initial state with $\phi_{a}\lr{0} = \xi_{a}$, $\frac{d}{dt} \, \phi_{a}\lr{0} = 0$ is an eigenstate of the Floquet matrix of the equation (\ref{second_order_eqm_phi_ev}) with eigenvalue $\lambda < 1$. Otherwise $\phi_{a}\lr{t}$ will also contain the mixture of the second independent solution of (\ref{second_order_eqm_phi_ev}), which grows exponentially. Therefore the time derivative of $\phi_{a}$ should be equal to zero at times $t = n T$: $\frac{d}{dt} \, \phi_{a}\lr{n T} = 0$. Since $x\lr{t}$ undergoes one-dimensional motion in a potential well $\tilde{V}\lr{x}$ with the initial condition $\frac{d}{dt} \, x\lr{0} = 0$, $x\lr{t}$ is symmetric under time reversal: $x\lr{t} = x\lr{n T  - t}$ for any $n$. The equation (\ref{second_order_eqm_phi_ev}) is therefore also invariant under time reversal, and if one evolves the system forward in time from $t = 0$ to $t = n T$ and back to $t = 0$, one should arrive at the initial value $\phi_{a}\lr{0}$. On the other hand, this value should be equal to $\lambda^{2 n} \phi_{a}\lr{0}$. We have thus to conclude that if $\lambda$ is real, it can only be equal to one, and the equation (\ref{second_order_eqm_x_ev}) has no exponentially decaying solutions. There can only be the solutions which oscillate periodically with time. For example, a typical solution of (\ref{second_order_eqm_phi_ev}) for the potential $V\lr{x} = \frac{w^{2} x}{2} + \frac{\lambda x^{2}}{4}$, $N V\lr{\frac{\phi_{a}^{2}}{N}} = \frac{w^{2} \phi_{a}^{2}}{2} + \frac{\lambda \lr{\phi_{a}^{2}}^{2} }{4 N}$ at the inverse temperature $\beta = 0.5$ is plotted on Fig. \ref{fig:phi_vs_t}. Dots denote numerical solution for $\phi_{a}\lr{t}$, which was obtained by numerically solving the saddle-point equations for $x\lr{t}$ and $\phi_{a}\lr{t}$ \cite{MakeenkoGaugeMethods, Feinberg:95:1, Fateev:81:1} on the $T$-shaped contour in the plane of complex time, as explained above. At not very large $t$ the amplitude of oscillations of $\phi_{a}\lr{t}$ indeed decreases, but at larger times $\phi_{a}\lr{t}$ oscillates in a beat-like manner. Note that such solution is completely different from the classical one, where both $x\lr{t}$ and $\phi_{a}\lr{t}$ oscillate with equal frequencies without any beats. The beats in $\phi_{a}\lr{t}$ are therefore a purely quantum phenomenon.

 It appears that the system described by the Hamiltonian (\ref{Hamiltonian}) does not have an asymptotic property of redistributing a perturbation of a small number of degrees of freedom over all configuration space at any finite temperature. Instead there is a continuous swapping of energy between an excited degree of freedom and all other degrees of freedom. This may be related to the fact that the system (\ref{Hamiltonian}) has as many integrals of motion (the components of angular momentum) as there are degrees of freedom. Therefore the interactions between all $N$ degrees of freedom should be arranged in some more complicated way in order to achieve ``scrambling''. Perhaps one should indeed give them a structure of large matrices, as suggested in \cite{Susskind:08:1, Polchinski:08:1}. However, the equations (\ref{second_order_eqm_phi_ev}) and (\ref{second_order_eqm_x_ev}), which describe the dynamics of $O\lr{N}$-symmetric large-$N$ model and not just the properties of its ground state, may be of interest by themselves. It might be interesting to find similar equations in higher-dimensional large-$N$ theories or in matrix models.

\begin{acknowledgments}
 The author is grateful to E. T. Akhmedov and M. I. Polikarpov for interesting discussions which motivated this work. This work was partly supported by Grants RFBR Nos. 06-02-04010-NNIO-a, 08-02-00661-a, 06-02-17012, and DFG-RFBR 436 RUS, BRFBR F08D-005, a grant for scientific schools No. NSh-679.2008.2, by the Federal Program of the Russian Ministry of Industry, Science and Technology No. 40.052.1.1.1112 and by the Russian Federal Agency for Nuclear Power.
\end{acknowledgments}


\end{document}